# DETECT KERNEL-MODE ROOTKITS VIA REAL TIME LOGGING & CONTROLLING MEMORY ACCESS

Satoshi Tanda

CrowdStrike, Inc

Vancouver, Canada

Igor Korkin

Independent Researcher

Moscow, Russia

{tanda.sat, igor.korkin}@gmail.com

## ABSTRACT

Modern malware and spyware platforms attack existing antivirus solutions and even Microsoft PatchGuard. To protect users and business systems new technologies developed by Intel and AMD CPUs may be applied. To deal with the new malware we propose monitoring and controlling access to the memory in real time using Intel VT-x with EPT. We have checked this concept by developing MemoryMonRWX, which is a bare-metal hypervisor. MemoryMonRWX is able to track and trap all types of memory access: read, write, and execute. MemoryMonRWX also has the following competitive advantages: fine-grained analysis, support of multi-core CPUs and 64-bit Windows 10. MemoryMonRWX is able to protect critical kernel memory areas even when PatchGuard has been disabled by malware. Its main innovative features are as follows: guaranteed interception of every memory access, resilience, and low performance degradation.

## 1. INTRODUCTION

Modern malware attacks on Windows machines are becoming increasingly sophisticated and extremely difficult to detect. Newest integrated security mechanisms on the modern Windows 10 x64 such as Kernel Mode Code Signing (KMCS) and Kernel Patch Protection (KPP) also known as PatchGuard are unable to prevent malicious activity.

Modern malware attacks are 'surgical' and infect networks of huge organizations even when their computers, have never been connected to the Internet – 'air-gapped' computers' (Paganini, 2014). Let us consider some recent incidents with the following malware: Turla rootkit, which remained undiscovered for at least three years and ProjectSauron, which has never been stored on a disk.

According to the security response by Symantec, Turla trojan which was created by the Waterbug hackers group successfully compromised more than 4,500 computers from 100 countries (Symantec, 2016). Even the Swiss Federal Department of Defense (GovCERT, 2016) was under a cyber-espionage attack via Turla (Paganini, 2016). This malware remained undiscovered for at least three years due to its stealth features, which helped to overcome both built-in security Windows and anti-virus signature based mechanisms. The authors of Turla rootkit proposed a new method to overcome Driver Signature Enforcement. A rootkit loads a legitimate signed driver and after that by using its vulnerability loads a malware driver. As a result, it defeats the Driver Signature Enforcement and makes it possible to load any kernel-mode driver even without any digital sign (G Data, 2014a; Rascagnères, 2016; Baranov, 2014). This malware hides its file system and registry activity by hooking the corresponding kernel-mode OS functions. To do this on a 64-bit system, malware bypasses PatchGuard without rebooting, which makes Windows kernel vulnerable to any manipulations again, such as Direct Kernel-mode Object Manipulation (DKOM) and hooking (G Data, 2014b).

AV expert from McAfee has demonstrated the ability of KPP-Destroyer utility to defeat PatchGuard on modern Windows 8.1 x64, which makes Windows kernel vulnerable to common well-known rootkit techniques. This tool has been used and improved by hackers (Intel, 2014; Rascagneres, 2015).

The authors underline that PatchGuard is vulnerable to kernel-level attacks because it is located in the same environment with rootkits (Yan, Luo, Feng, Pan, & Safi, 2015). The TDL4/TDSS rootkit family disables PatchGuard by modifying the system's boot loader.

Another malware platform for cyber espionage was identified by Symantec and Kaspersky Lab as ProjectSauron and according to their reports this malware had eluded security researchers for at least five years (Dockrill, 2016). This malware was deliberately created to confuse AV experts and to prevent its analysis. To achieve this, the indicators of compromise or patterns, which are normally used by AV experts, were removed. ProjectSauron also resides only in the computer memory without saving itself to the hard disk drive, which renders existing AV techniques pointless (Baranov, 2016).

According to the paper (Prakash, Venkataramani, Yin, & Lin, 2015) "a kernel rootkit, can often tamper with kernel memory data, putting the trustworthiness of memory analysis under question." These authors state "moreover, while it is widely accepted that value manipulation attacks pose a threat to memory analysis, its severity has not been explored and well understood."

These authors proposed improving the DKOM attack that targets the OS scheduler. They also showed that it cannot be detected by any of the existing techniques (Graziano, Flore, Lanzi, & Balzarotti, 2016; Graziano, 2016).

Detection of malicious binaries with digital certificates is becoming increasingly difficult. Cyber security researchers keep sharing new techniques to overcome Windows security mechanisms (KMCS) in the recent Black Hat USA 2016 conference (Nipravsky, 2016). The idea of infection of digitally signed files without altering hashes was based on inserting a payload code into the header attribute certification table. Because Windows excludes this field from the hash calculations, the file certificate remains valid. According to the recent McAfee Labs Threats Report (McAfee, 2016) the total number of malicious signed binaries increased by 3 million during the first 6 months of 2016.

Experts from Kaspersky Lab have published the newest set of malware tricks, which make it difficult to reveal malware (Bartholomew & Guerrero-Saade, 2016).

The authors Jadhav, Vidyarthi, & Hemavathy (2016) prove that modern malware are prepared thoroughly enough to prevent their detection even by high skilled AV experts. Hackers "leave no signature, and so they never get caught. This happens due to the absence of signature or behavior information in the security systems." At the same time, we are able to detect this new unknown malware because "in many cases evasive behaviors can be used as a signal for evasive malware detection."

Thus existing protection approaches of computer systems are no longer working. Driver Signature Enforcement cannot prevent installation of signed malware, PatchGuard is not resilient to malware counter-measures, modern AV products are unable to detect malware even for several years.

The purpose of this paper is to present the design, implementation, and evaluation of a new hypervisor-based system that reliably provides privacy and integrity of memory data as well as giving behavior information on memory access in real time. To detect unauthorized memory access, we propose a new memory monitor system – MemoryMonRWX, which has the capability to track all memory accesses.

**Thread model**. We will consider the following basic scenarios of malware attacks in the kernel-mode:

1. Stealing sensitive data, such as crypto keys and private users' data.
2. Manipulation with memory content, such as hooking, unlinking, and patching.
3. Execution of unknown code fragments.

**Scenario 1**. Malware reads the sensitive data from memory, such as private users' data, cryptographic key, passwords, hashes, data and code of 3[rd] party drivers. Recent research papers show the advance and importance of this topic. The way of extracting crypto keys from BitLocker is presented here (White, 2015). Thorough analysis of TrueCrypt utility and ways to retrieve user's crypto keys are presented by Baluda et al. (2015). Security analysis of BestCrypt was carried out by Souček (2016), the data leaks issues has been revealed.

Moreover, kernel mode exploits usually read Windows kernel internals data, for example HalDispatchTable (Cardona, 2017). Hence there is a need to manage the access to this data as well.

**Scenario 2**. Malware disables PatchGuard and illegally modifies the critical parts of system memory. Malware hooks functions by tampering System Service Descriptor Table (SSDT), hides OS-objects, such as process and drivers, by unlinking and patching corresponding structures from lists Sim & Lee (2016); Li, Wu, & Liu (2016). As a result, this involves memory modification of no less than 8 bytes for 64-bit OS. Rootkit can further protect this unlinked structure by overwriting its fields. So, this means no less than one-byte data modifications (Haruyama & Suzuki, 2012).

Malware can also hijack the kernel control transfers by Kernel Object Hooking (KOH), including the violation of control-flow integrity. For example, changing JZ to JNZ modifies one byte of code (Wang & Guo, 2016).

**Scenario 3**. Malware deletes or modifies all information about itself from the system. As a result, there are only executable code fragments in the memory, which do not belong to any of the registered drivers. This idea was originally proposed by Korkin & Nesterow, 2016.

To process all these scenarios for attackers, we propose the following logging scenarios. The visualization of malware attacks examples and the registered output are in the Figure 1 and Table 1.

**Logging Scenario 1**. SyspiciousDriver.sys tries to steal sensitive data. To achieve this its code block, which is loaded to address 'SourceAddr1', reads the memory data, which is located on the address 'DestinationAddr1'. As a result, the output needs to register the following triple: 'SourceAddr1 – Read – DestinationAddr1'.

**Logging Scenario 2**. The SuspiciousDrv.sys tries to hook a system table function. In this situation its code block, which is loaded to the address 'SourceAddr2', writes to the memory fragment, which is located on the address 'DestinationAddr2'. After this, the output will include the following items: 'SourceAddr2 – Write – DestinationAddr2'.

**Logging Scenario 3**. The HiddenDrv.sys hides itself by deliberately deleting all related information from the system lists. As a result, we have only executable code, which is loaded on the 'SourceAddr3' in the kernel-mode memory. In order to detect it, the output needs to add the following entry: 'SourceAddr3 – Execute – SourceAddr3'.

Table 1 Example of preliminary output for revealing malware attacks

| # | Source address | Access Type | Destination address |
|---|---|---|---|
| 1 | SourceAddr1 | Read | DestinationAddr1 |
| 2 | SourceAddr2 | Write | DestinationAddr2 |
| 3 | SourceAddr3 | Exe-cute | SourceAddr3 |
| … | … | … | … |

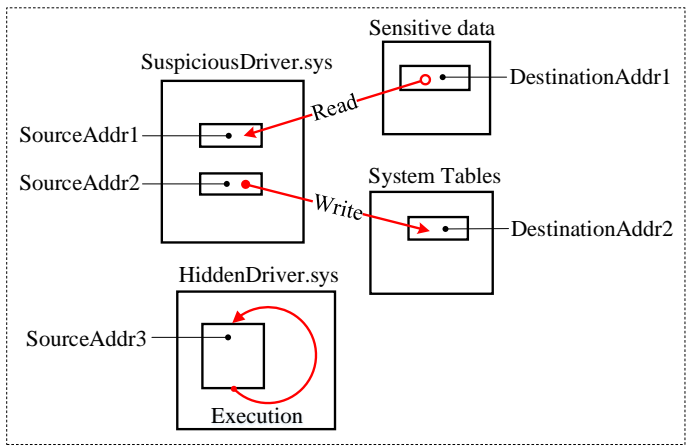

Figure 1 Examples of malware attacks in memory and the proposed log of the detection system

The following is required for solving the task:

- for each memory access attempt, we need to log the following three items: source address, destination address, and type of accessing – read, write, or execute;
- we need to specify two intervals: one for the source addresses and one for the destination addresses. The accesses from only these two intervals will be tracked;
- the interval of addresses needs to support two values – start and finish addresses as well as only one value – a fixed virtual address;
- the prototype needs to support the modern Windows 10 x64 and multi-core CPU.

For the proposed memory monitoring system, we restrict CPU requirements with Nehalem microarchitecture (Wikiwand, n.d.), which supports both technologies Intel VT-x and Intel VT-x with EPT.

This paper is in four sections. Section 2 focuses on the comparative analysis of the existing ways for logging memory access. In the first part we will analyze methods, which work inside an operating system: tracking memory management routines and the methods based on replacing page fault manager. The second part covers the analysis of hypervisor-based methods for tracking memory access. We provide a review of other recent papers and their ideas. At the end of this section we select the possible avenues for further development.

The design of the proposed system MemoryMonRWX is presented in the third section of this paper. We describe the architecture and major components of this system. The details of interactions of major components in three cases on controlling read, write, and execute access are provided. To outline the advantages of MemoryMonRWX we present three demos: integrity case, confidentiality case, and an example of the analysis of real rootkit. We evaluate the benchmarks of MemoryMonRWX and demonstrate that the degradation of system performance is about 10%.

Section 4 contains the main conclusions and further research directions.

## 2. BACKGROUND

In this section related papers are reviewed as well as existing prototypes according to the requirements previously mentioned. There are several hardware based solutions which are able to monitor memory access using FPGA programmable platform (Morgan et al., 2015; Lee, et al., 2013). These approaches are only applicable in the laboratory situation, because it is hard to distribute and upgrade them; so they will be omitted and instead the focus will be on software-based methods.

All software methods for monitoring memory access can be divided into two groups: first those based on operating system facilities and second those based on hardware virtualization technology – otherwise known as OS-based and hypervisor-based, correspondingly (Bauman, et al., 2015). The classification of these methods is presented in Figure 2.

OS-based methods can be sub-divided into two subgroups. The first subgroup monitors memory access by tracking calls of memory management functions, while the second one applies handling page fault exception (#PF) by the Interrupt Descriptor Table (IDT) inside the OS.

Hypervisor-based methods can be divided into the two subgroups according the technologies, which they are based on. The first subgroup leverages hypervisor facilities to handle page fault exception, the second subgroup applies new Intel VT-x with Extended Page Tables (EPT) technology to track memory access. The proposed MemoryMonRWX system is based on the EPT technology.

Next all software methods will be analyzed and we will discover the most reliable and resistant method.

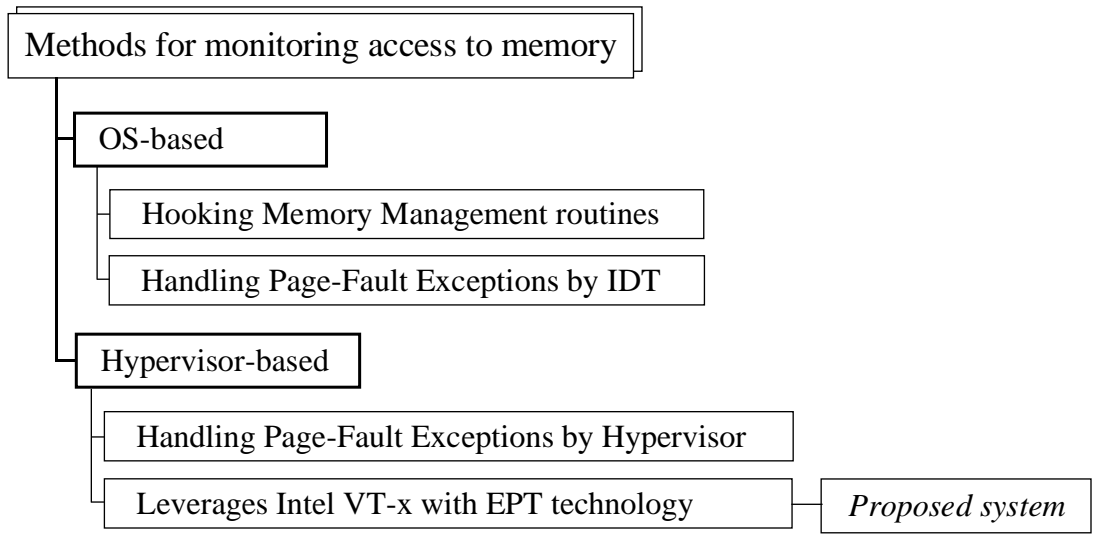

Figure 2 Classification of methods for monitoring access to memory

## 2.1. OS-based Methods

In this section we will analyze methods, which work inside a Windows operation system and do not require any specific CPU support. Initially the focus will be on applying tracking memory management routines to monitor access to the memory. Next, we will deal with tracking memory access via marking pages as non-present and replacing the page fault handler.

### 2.1.1. Hooking Memory Management Routines

During the lifecycle of a kernel-mode driver several kernel mode objects or structures will have been added into the memory. During driver's installation the corresponding structures are added into the system lists (Mayer, 2015); also a driver can allocate memory for its own purposes – all these manipulations can be tracked and used as a source to detect a malicious driver.

One of the ways used to implement hooking functions is the rewriting of an address of memory allocation routine and applying trampoline function.

The idea of monitoring the execution of an operating system and tracking the newly created kernel mode objects was proposed by Prakash et al., (2015). The authors suggested hooking memory allocation and deallocation routines in the kernel: ExAllocatePoolWithTag, ExFreePoolWithTag and MmLoadSystemImage from ntoskrnl.exe. Their ideas have been tested on 32-bit Windows XP with Service Pack 3 and Windows 7.

A similar idea of capturing kernel mode object allocation and deallocation events to dynamically identify kernel-mode objects was also proposed by Rhee et al., (2010). The authors considered two basic types of malware attack: privilege escalation using direct memory manipulation and dynamic kernel mode object hooking. They emphasize that performance is not a primary concern for their prototype, which is designed for use in non-production scenarios such as honeypot monitoring, etc.

The method of intercepting kernel-mode functions by using inline hooks in a stable manner on multi-cores processor systems was proposed by Milković (2012).

It is also possible to hook kernel-mode functions by applying well-known techniques of modifying pointer values onto the System Services Dispatching Table (Matrosov, Rodionov, & Bratus, 2016) and using the Stealth Hook technique and Redirector Stubs to conceal hooking (Ligh et al., 2014).

All these hooking approaches work well only on 32-bit Windows OSes, while the more popular 64-bit Windows include built-in Kernel Patch

Protection or PatchGuard. According to the blog PatchGuard "is intended to protect critical kernel structures from being easily modified from unauthorized entities" (Block, 2015). PatchGuard also controls the integrity of Windows kernel, including ntoskrnl.exe (Irfan et al., 2013; Comodo, 2013).

### 2.1.2. Handling Page Fault Exceptions by IDT

This method is based on memory mapping on Intel x86 in protected mode. The idea of intercepting memory mapping process using IDT for rootkit purposes was first presented by Sparks & Butler (2005). Below we will discuss the details of memory mapping process and how to apply them to monitor memory access. We will provide three scenarios of trapping memory access and also cover with the disadvantages of this methodology.

The process of memory mapping or memory paging is explained by Intel (2016) and includes the following phases. When a memory access to the page occurs, a CPU starts page table walk to find the physical address. CPU then checks the access type by reading corresponding Page Table Entry (PTE) status bits. If the page is valid (meaning that its bits are set) and there is no conflict with the access type, the CPU then calculates the corresponding physical address of the page, using the page frame number (PFN) from this PTE.

This is a frequent scenario. However, the access violation case is also possible: according to the Windows source code "the access fault was detected due to either an access violation, a PTE with the present bit clear, or a valid PTE with the dirty bit clear and a write operation" (Microsoft, n.d.-a). In this situation after checking PTE bits, the CPU raises a page fault exception (#PF). Following this the control goes to the page fault handler code, whose address is located in the IDT. The example of the source code of the page fault handler code is presented in the function nt!KiTrap0E within the file Microsoft (n.d.-b). This code processes all the required work for loading memory pages, configuring PTEs and continue control to the OS.

Sparks and Butler proposed hiding of the memory page by deliberately marking corresponding PTE as non-present and also by replacing the page fault handler code, which helps to differentiate page view. This method can be applied to monitoring memory access as well. Figure 3 shows the principles of tracking memory access.

Let us consider the case of secret data protection from unauthorized access. Secret data is located on page C. To do this we change the corresponding page table entry by clearing the Present bit (P bit). Once an unknown driver 'Drv.sys', has been loaded on page A and page B, and tries to read the secret data, CPU starts memory translation to retrieve the content of page C. To achieve this CPU reads the Page C PTE and checks if the result is in conflict with the access type. In our case we have access violation: 'Drv.sys' reads a page with zero present bit and CPU raises a #PF exception. CPU processes this #PF by passing control to the code of page fault handler via IDT, which stores a link to its code. We can then replace the original page fault handler or its code and add a new processing algorithm. In the page fault handler code, we can receive the saved instruction pointer (SourceAddr), faulting address (DestinationAddr) and with this information we can realize various processing algorithms.

We will consider the following three scenarios of page fault handler code.

**Scenario 1. Protecting secret data from being read.** To protect secret data from unauthorized reading we clear P bit in the secret page PTE. During reading from this page the #PF (page not present fault) will be raised and page fault handler code starts to go (Eranian & Mosberger, 2002). We can update the page fault handler algorithm to filter this access violation in the appropriate way using the saved instruction pointer as SourceAddr and faulting address as DestinationAddr. As a result, we are able to return the 'fake' page to the caller.

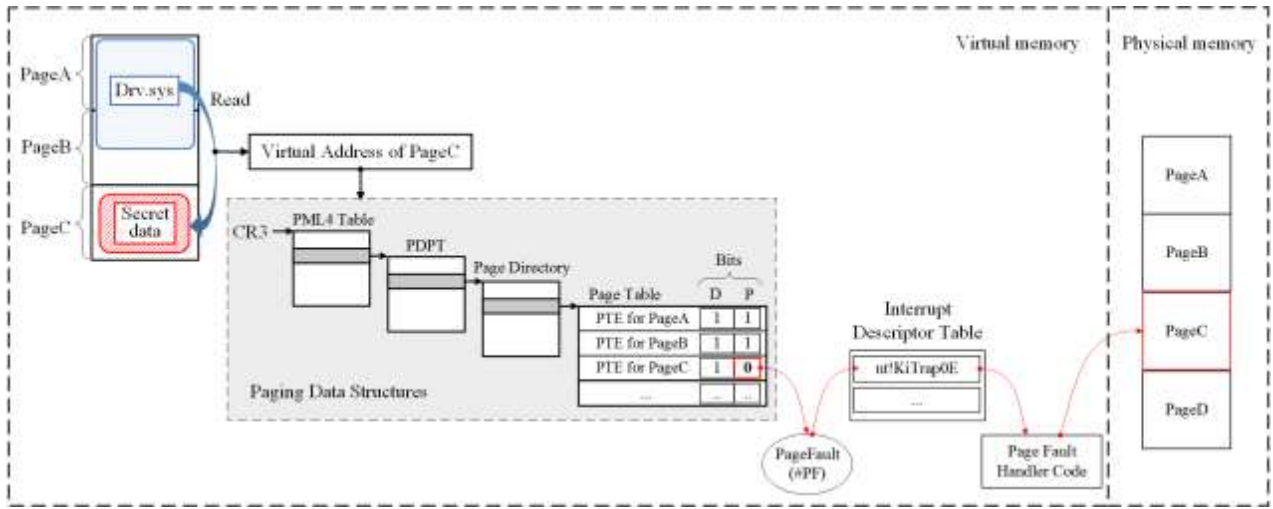

Figure 3 Log and manage access to memory pages by setting corresponding page table entry as non-present and replacing the page fault handler

**Scenario 2. Protecting system data from being modified.** To protect the memory pages from unauthorized modification, for example by providing the integrity of the system structures, we clear Dirty bit (D bit) in the PTE, which will correspond to the page with the system structures. During any writing access to this page, CPU will raise a #PF (dirty-bit fault). Using a similar pathway from scenario 1 we can update the page fault handler to process this violation in an appropriate manner.

**Scenario 3. Detecting unauthorized code execution.** Trapping execution appears to be a more complicated task, because for IA-32 architecture there is no way to distinguish the execute accesses from read and write. To reveal execute accesses we use manipulation from Scenario 1, which is applied, adapting the proposals of Sparks & Butler (2005). An execute access is achieved if SourceAddr and DestinationAddr are equal.

Existing approaches, which mark pages as non-present and replace the page fault handler can be analyzed.

The idea to control access to the pages by trapping page fault exceptions was proposed by Backes et al. (2014). This group authors attempted to avoid code reuse attacks, and this approach prevents programs from reading executable memory.

Another system (Xu et al., 2015) focuses on controlled-channel attacks, which extract sensitive information from the application. This attack is based on restriction access to the particular code or the data page by editing the page table attributes directly. When the application tries to access one of these pages, a page fault will occur. When a page fault happens, the authors system will log the page fault event, and enable access to the page and remove access from the previous page. Their system records full byte-granular page fault traces of both code and data pages.

The idea of monitoring memory access by page level tracking is used in the Omnipack kernel driver to detect when the program has removed the various layers of packing. Omnipack tracks written and written-then-executed memory pages. This system enforces a write-xor-execute policy (W+X) on the memory pages of the suspicious program to detect any attempts to execute the generated code during unpacking. A similar idea of W+X protection policy is also used in SecVisor (Seshadri et al., 2007).

The method of process' address space protection via the mechanism of intercepting each time the processor asserts the page fault interrupt to signal the access violation was used in the KLIMAX (Stefano, Cristiano, & Bruno, 2011).

The SPIDER system by Deng, Zhang, & Xu (2013) realizes data watch point and enables monitoring memory to read/write at any address. These workers underline two limitations of applying existing techniques based on the page-level mechanism for trapping execution. First, every instruction for fetching or data access in the non-present page will cause a page fault. This would result in a prohibitively high performance overhead. Second, the modified page table and page fault handler could still be detected by kernel-mode drivers.

However, this method of trapping memory access by using PTE modification and replacing page fault handler code has several weaknesses. Sparks and Butler (2005) have shown that this method does not support 4 megabyte pages and, moreover, a replaced page fault handler can be easily revealed and this can help to detect this method. Due to the fact that page fault handler code is an intermediary, memory monitoring will have a major impact on the system performance. In addition, there are issues of porting this method to the CPUs with multiple cores (Priyadarshi, 2016).

The main disadvantage of OS-based methods are as follows: they can be easily detected and disabled by kernel-mode malware. The hypervisor-based methods are relatively stealthier and more resilient, but they require a CPU with hardware virtualization support.

## 2.2. Hypervisor-based Methods

Methods described in this section require hardware virtualization processor features, which are enabled in all modern CPUs. The first method uses Intel Virtualization Technology (VT-x) without any specific features. As a result, this method will work even on the legacy Intel Core 2 CPU. The second method leverages Intel VT-x with EPT technology, which can be used with 2nd generation of Intel – family i3, i5, and i7. One of the recently analyzed papers requires processors with support from Intel Processor Trace (PT) technology, which is integrated only in newest CPUs beginning with 5th generation.

### 2.2.1. Handling Page-Fault Exceptions by Hypervisor

This method leverages hardware virtualization technology into monitoring access to memory by processing the page fault exception. This method, like the previous one, modifies the page table entries or the attributes of the memory pages, access which should be controlled. Any access to this page will generate the #PF and cause VM-exit, which will be handled by the hypervisor, see Figure 4.

To set up the hypervisor for processing #PF we need to configure Virtual Machine Control Structure (VMCS). This can be achieved by setting the 14th bit in the Exception Bitmap from VMCS->VM-execution control fields.

Applying this method, the hypervisor is able to catch both SourceAddr and DestinationAddr addresses, realizing various security scenarios. According to the page, which reveals illegal memory access: a hypervisor gets the address of the trapped instruction from EIP (Cheng, Ding, & Deng, 2013). Some recent examples of this method will be given and finally the drawbacks of the method will be presented.

Kuniyasu et al. (2014) proposed the DriverGuard hypervisor to protect industrial infrastructure systems from Advanced Persistent Threat (APT). The authors considered, that most of these threats "are zero-day attacks and signature based security tools cannot detect these attacks." Their hypervisor "prevents malicious write-access to code region that causes Blue Screen of Death of Windows, and malicious read- and write- access to data region which causes information leakage." DriverGuard manages PTE and changes the Present bit (P bit). As a result, all access to the page causes a page fault, which is hooked by DriverGuard; it analyzes whether the access comes from a legitimate code or not. If a legitimate code accesses the memory, DriverGuard will apply a new stealth breakpoint technique using hardware breakpoints in the single step mode. It enables single step mode by setting Monitor Trap Flag (MTF) bit in the VMCS. DriverGuard recognizes the memory region with sensitive data using "tag" value. Memory regions which are allocated dynamically by ExAllocatePoolWithTag with this "tag" value will be protected. Hackers can reveal this "tag" value and use the same tag in their malware. The authors admitted that page fault is slower than software interrupt and "it will make performance degradation."

Another protection system – MOSKG, which is countering kernel rootkits with a secure paging mechanism was presented by Yan, Luo, Feng, Pan, & Safi (2015e). The primary goal of this paper is to prevent rootkits by preserving critical kernel mode data from being manipulated by DKOM and page mapping attacks. These authors underline the main challenges as "the dynamic data can be modified legally by the OS or illegally by using the rootkits, but we have to distinguish the legal operations from the illegal ones." To validate the legitimacy of write operation to dynamic data and page mapping operations they make use of the shadow page tables (SPTs) in the hypervisor to mark the machine page, which in turn contains the protected data as read only. As a result, "whenever an instruction attempts to write the marked page, the page fault handler in the hypervisor will be called." They underline the limitations of their solution, one of which is "that the extent of protection is not sufficient." The next limitation is that "the rootkits may seek out other unprotected data to compromise the target OS. These attacks might be able to circumvent some portions of MOSKG architecture."

Wang & Jiang (2010) consider the issue of hypervisor integrity protection. They assume there is a threat model in which attackers are able to exploit software vulnerabilities to overwrite any memory data. They focused on the hypervisors and that "in current hypervisors (e.g., Xen and KVM) and OS kernels (e.g. Windows and Linux), their page tables are all writable." Experiments have shown that modification of "even one bit in a page table entry could well be enough to subvert the entire protection." The authors proposed HyperSafe, a lightweight approach, which protects the hypervisor's code and data from being compromised. To provide the W+X-based integrity HyperSafe marks the page tables as read-only and turns on the Write Protect bit (WP) in the register CR0. This bit controls the way a hypervisor code interacts with the write protection bits. As a result, any write attempts to modify them at runtime will be trapped by the hypervisor. HyperSafe is able to protect only open source hypervisor. The support of closed source 3[rd] party drivers still remains a major challenge.

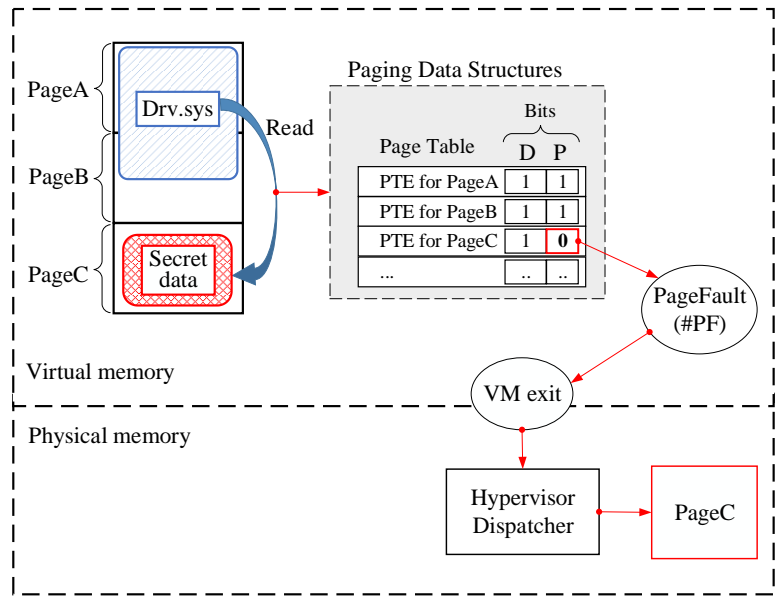

Figure 4 Controlling access to the guest OS virtual memory via marking corresponding page table entry as a non-present and handling page fault exceptions by hypervisor

Patents exist which also cover memory tracking ideas by marking guest kernel page table entries as not present and controlling page fault. A patent by Dang, Mohinder, & Srivastava (2015) proposes a hypervisor's page fault handler, which may conditionally allow or deny access to or execution of the appropriate guest kernel pages. Using the fact that the assignee of this patent is McAfee, Inc. from Santa Clara, California (USA), we can infer that this methodology is used in the McAfee Antivirus or its internal products. In another patent (Traut, Hendel, & Vega, 2007) the processing interrupts to maintain the modified flags of the page table entries and this is a significant source of the slowdown for a shadow page table implementation.

Page fault protection mechanism is used for monitoring the access to kernel-mode pages but also to user-mode pages. U-HIPE is the prototype for user-mode memory protection, which injects a page fault in the guest VM (Luțaș, et al., 2016).

Srivastava & Giffin (2011) explore the idea of monitoring untrusted kernel-mode execution by separating page tables for data and for drivers. This separation forces all control flows spanning the kernel-driver interface to induce page faults which are then handled by the code in the hypervisor and this verifies the legitimacy of the control flow. As a result, a hypervisor-based system called Gateway was created. This system traces the behavior of kernel malware by monitoring kernel APIs functions invoked by drivers.

A similar method for processing page faults was used in the hypervisor-based system HyperSleuth for tracing system calls (Martignoniey al 2010). Because all system calls invocations go through a common gate, whose address is defined by SYSENTER_EIP register, these workers shadow the values of this register and the value of the shadow copy to the address of a non-existent memory location. Afterwards, all system calls invocations result in a page fault exception. As a result, HyperSleuth traps and saves all system calls to the log, which is then transmitted via the network to the trusted host.

Another idea is trapping system calls using a virtual machine introspection mechanism (Pfoh, et al., 2011). This Nitro system works with the following system calls: user defined interruption; and SYSCALL / SYSENTER instructions. This system is not able to monitor function calls in the kernel-mode.

Azab et al. (2009) present a hypervisor-based system that measures the integrity of virtual machines – HIMA. The idea of protecting guest memory using page access permissions was also used in the HIMA. This author's system applies facilities of No eXecute bit (NX bit) of a page table entry. If this bit is 1, the page is assumed to only retain data. Any instruction execution on this page will cause a page fault exception, which will be trapped by the hypervisor. Moreover, the authors proposed to prevent programs from marking executable pages as non-writable, which provides trapping of any modification of the memory pages.

This memory trapping method is also used in PhD research. Thus Yan (2013) deals with malware analysis by virtualization and demonstrates that this memory monitoring method is not resilient for Denial-of-Service (DoS) attack. The problem is that malware can induce a large number of page faults exceptions, and each of them involves an exit to the hypervisor. This exception flood launches a DoS attack on the recorded log and renders its further analysis difficult.

As well as this DoS vulnerability this method has the following drawbacks:

- It is not stealthy: the modification of page attributes or PTE.flags is visible from a guest and can be easily revealed by malware;
- It is not lightweight: each page fault will take some time for processing and in real time will result in significant performance overheads;
- It does not fully support multi-core CPU: since a PTE exists only for a page and is shared by all cores, its modification it affects other processors' contexts as well.

In the next section we will present other methods, which exclude these drawbacks, but require CPU with VT-x and EPT support.

### 2.2.2. Leverages Intel Extended Page Tables technology

This section covers Intel VT-x with Extended Page Tables (EPT) technology, which is a new feature of hardware virtualization. We present the details of how EPT mechanism is working and the ways of leveraging EPT for tracking and trapping access to the memory.

**New hardware virtualization feature – EPT is the source of inspiration for monitoring memory access.**

There are two serious drawbacks of Intel VT-x technology presented in the previous section. Firstly, there is hypervisor performance overhead associated with memory management and secondly the size of guest physical memory is limited by host physical memory. The idea of Second Layer Address Translation (SLAT) or Two-Dimensional Paging has been designed to reduce the memory and power overhead costs through hardware optimization of the page table management.

The SLAT technology has been integrated in the Intel CPUs since Nehalem microarchitecture in the first Core i3, i5, and i7. In Intel terminology this technology is 'Intel VT-x with Extended Page Tables (EPT)'. Similar technology has been issued by AMD and this is called 'Nested Page Table (NPT)'. In this review the focus is on EPT implementation in the Intel CPUs, but the findings apply more generally as well.

We will show how the EPT mechanism works and how it can be used for monitoring memory. EPT technology helps to virtualize guest physical memory and as a result enhances CPU facilities using paging data structures also known as 'EPT layout'. The algorithm of the EPT data structures, which translates the guest physical address to the host physical address, is similar to the familiar algorithm of paging structures in the protected mode, which translates the guest virtual address to the guest physical address. The content and organization of EPT paging structures are analogous to the paging structure in the protected mode or x86-64 page tables. There are however several differences between the content of EPT and guest paging structures (Grehan, 2014).

EPT paging structures include the following tables: Page Map Level 4 (EPT PML4), Page-Directory-Pointer Table (EPT PDPT), Page Directory (EPT PD), and Page Table (EPT PT), as shown in Figure 5. Hypervisor needs to allocate memory for all these tables and place their content. Using differing configurations of EPT structures the hypervisor can provide various memory paging scenarios.

In this paper we will consider a simple scenario with 'memory 1:1 mapping', which translates guest physical address into the same physical address (Uty, & Saman, 2016).

During each memory access inside guest operating system (guest OS), initially the guest paging data structures are involved. Finally, the EPT structures are to convert the received guest physical address into the host physical address.

We can intercept memory access to the page by modifying the bits in the corresponding entry in the EPT Page table, while the entries in other tables EPT PML4, EPT PDPT, and EPT PD have their own default values.

EPT Page Table entry provides bits, which allow or disallow access to the corresponding page:

- bit#0 – "Read Access", indicates whether reads are allowed from the 4-KByte page;
- bit#1 – "Write Access", indicates whether writes are allowed from the 4-KByte page;
- bit#2 – "Execute Access", shows whether instruction fetches are allowed from the 4-KByte page.

According to the Intel manual (Intel, 2016) – 'Any attempts at disallowed accesses will involve EPT violation and will cause VM exits'. Hypervisor intercepts each EPT violation (VM Exit) and can implement specific algorithms, which help to provide cyber security as well as hiding malware data in the memory.

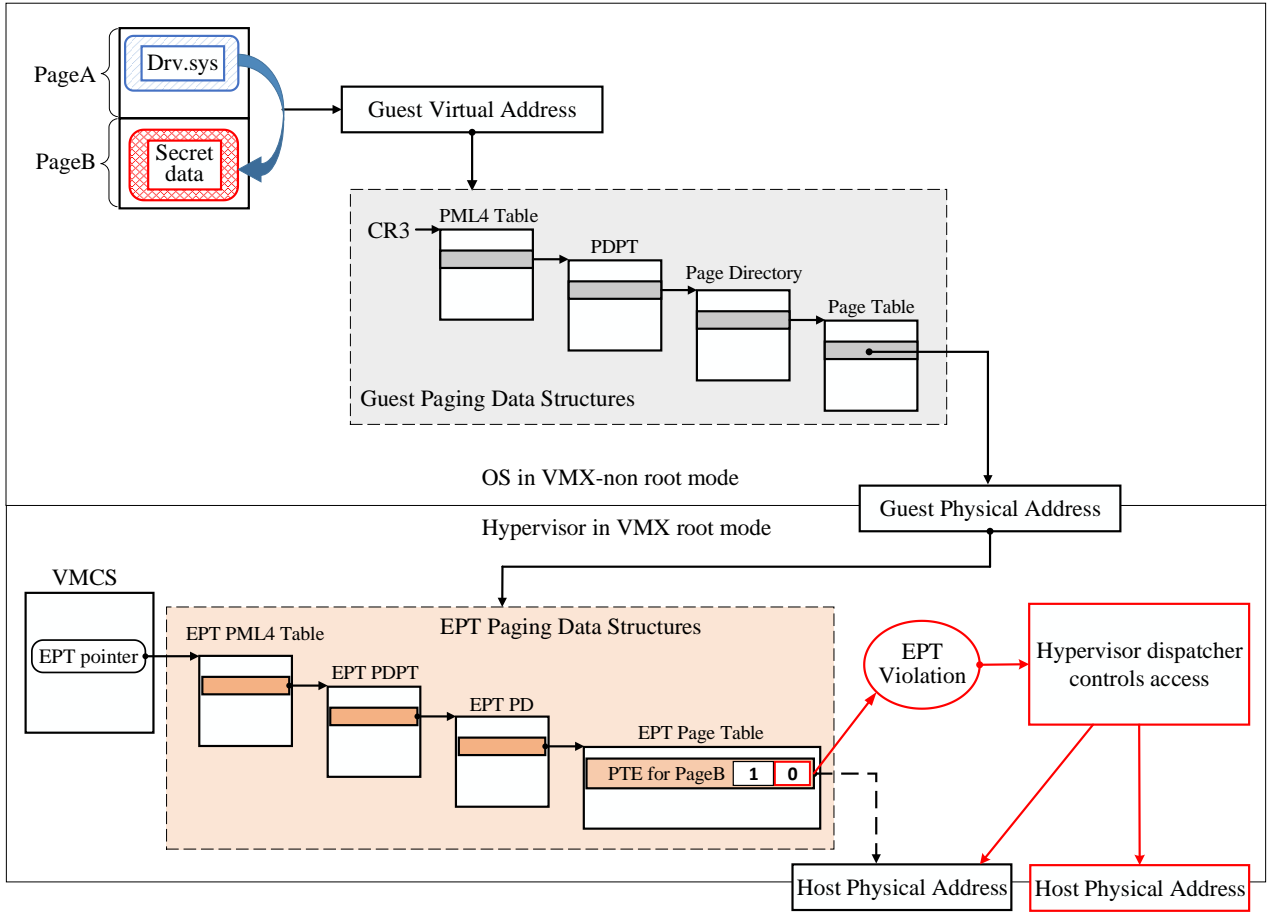

Figure 5 Controlling access to the guest OS virtual memory via using second layer address translation

Hypervisor can log any read and write memory access of the malware driver by resetting the first two bits of its EPT PT entry. It is able to protect the privacy of its memory page content by resetting bit#0 of its EPT PT entry and replacing the required physical page to another null page where the access is disallowed. Hypervisor is also able to provide the integrity of memory page by resetting bit#1 and trapping each write access to the page. Rootkit hypervisor can protect the malicious driver from antiviruses and Windows built-in security mechanisms such as PatchGuard. For example, to protect secret data on Page B from reading by Drv.sys, which is loaded on Page A we need to modify the bit 0 in EPT PT entry, which corresponds to the page B.

In the next section we review the papers, which leverage EPT technology, and also comment on their drawbacks.

**Analysis of EPT-based cyber security solutions**

First, the dynamic analysis system – DRAKVUF is able to track execution and tackle DKOM attacks (Lengyel et al., 2014). The DRAKVUF system uses VT-x and EPT technologies and is built on Xen hypervisor and the LibVMI library. The core technique is based on writing the opcode 0xCC at the code location deemed of interest. This manipulation is named as breakpoint injection and is trapped by the DRAKVUF hypervisor. This technique is able to automate the execution tracking of the entire OS and can trap all kernel functions. The breakpoint injection technique is protected by EPT page permission and enables an active virtual machine introspection. DRAKVUF adopted a novel approach to tackle DKOM attacks. To locate internal kernel structures DRAKVUF traps kernel heap allocations directly by using breakpoint injection for Windows functions, which are responsible for allocating memory for structures: ExAllocatePoolWithTag and ObCreateObject. This system detects the locating of all kernel structures by dynamically extracting the return address from the stack. Another interesting feature of DRAKVUF system is to monitor access to the file by tracking access to the corresponding _FILE_OBJECTs structure in the kernel-mode heap. DRAKVUF marks the page on which the structure is allocated as non-writable using EPT technology. This monitoring systems supports only Windows 7 in both 32 and 64-bit versions (Drakvuf, 2016).

In a second monitoring system CXPInspector by Willems, Hund, & Holz (2012) focuses on the analyzing of the state of a virtual machine from the outside. This system is based on the concept of Currently eXecutable Pages (CXP), which has the capability to observe the behavior of a program or even a complete OS and has two main uses. Firstly, it helps to analyze the behavior of kernel-mode malware and, secondly, it provides a performance profile of a single program or a whole OS. The authors leverage the EPT technology by virtualizing the memory management unit, and guarantees address space separation and hence it no longer requires hooking the page fault handler. The authors proposed an innovative approach of trapping access to the memory pages, which has instead been delegated out to the disk. To handle this situation CXPInspector injects a page fault into the guest OS, and this in turn forces the guest's page fault handler to page-in the required memory. It uses a single step mode by setting Trap flag (TF) in the EFLAGS register. CXPInspector is based exclusively on the use of EPT technology for tracing instruction fetches. CXPInspector monitors execution code by marking certain pages as non-executable. CXPInpector enriches generated logs with detailed information about the called functions and their argument values. The called function names are retrieved through the use of debug symbol information. The key feature of CXPInspector is its ability to record the memory addresses from which call/return originated. This becomes possible through the use of Intel Processor Trace (Intel PT) technology, which has been integrated into Intel CPU since their [5th] generation. CXPInspector is implemented on KVM hypervisor, which provides an interface to the QEMU toolset for 64 bit machines and Windows 7.

The next paper by Pham et al., (2014) proposed the design of a HyperTap, a hypervisor-level framework, which monitors a variety of system events and states. HyperTap is able to trap context switching, syscalls, instruction execution and memory accesses. HyperTap can be adapted for a wide range of reliability and security (RnS) policies. HyperTap protection algorithms are based on setting memory protection for the allocated Memory Mapped I/O area so that access to this area will trigger EPT_VIOLATION events. We focus on two main examples of applying HyperTap: (1) Hidden RootKit Detection (HRKD) and (2) Privilege Escalation Detection (PED). In order to detect a hidden user process or thread, HyperTap tracks thread switches by setting memory access permissions. The HRKD sets all memory pages that contain TSS structure as write-protected. As a result, each thread switch modifies the task state segment (TSS) structure, which is then rerouted to the hypervisor using EPT_VIOLATION. The algorithm of PED is based on the Ninja privilege detection systems. PED applies OS-level Ninja's checking rules whilst screening for unauthorized access. The idea behind this is to intercept fast system calls by using execute-protection on the system call entry point so that a guest attempt to execute this system call will generate EPT_VIOLATION. HyperTap is based on the KVM hypervisor and Linux kernel. These protection mechanisms have been tested on Windows XP, Vista, and 7.

The idea of applying hypervisors facilities and EPT technology for rootkit purposes has been reported by Uty & Saman, (2016). These authors focus on the invisible inline hooks aimed at the modification of the code section. Hypervisor maps guest physical address to the host physical addresses by using 1:1 mapping. The idea of using invisible inline hooks is based on copying physical aspects and the original page. This memory page copy evades any detection of integrity violation. Thus in order to hook the Windows internal function the authors inject 0xCC (breakpoint, #BP) into the original page, while the duplicated page becomes unpatched. The memory page with this inline hook therefore will have an executable page permission. Any read operation to this page will trigger the EPT violation and during its processing the hypervisor changes the page's mapping to the shadow one with read-write permissions. To speed up the authors proposed the use of two slightly different EPT structures and then simply switch between them. Using the idea of invisible inline hooks, the author has realized the keylogger and bypassed PatchGuard. In the first case authors hooked the KeyboardClassServiceCallBack routine and in the second case to hide a process they hooked to NtQuerySystemInformation. These hypervisor-based rootkits facilities have been successfully tested on Windows 7 x64 and Windows 8.1 x64.

We can see that there are the following drawbacks of these EPT-based studies:

- EPT technology operates on page granularity level, which is why all these studies proposed a page-level control, without fine-grained analysis;
- There is no solution which is able to monitor and control all possible memory access simultaneously: read, write, and execute;
- All analyzed security solutions are not flexible because they are based on huge platforms, such as Xen, KVM, QEMU etc.

We can conclude that EPT technology provides a huge opportunity to monitor and control access to the memory pages, and this can be used as the basis for the proposed solution.

### 2.3. Conclusion

The above analysis shows that the existing memory monitoring methods have the following drawbacks:

1. Both OS-based methods are vulnerable to kernel-mode malware manipulations.
2. Methods based on handling page fault violations via bare-metal hypervisor are not lightweight and will not support multi-core CPUs properly.
3. EPT-based methods provide neither a fine-grained analysis nor the ability to trap all memory access.

The summary with the comparison analysis of the major papers and projects is given in the Table 2, where $2^{nd}$ generation CPU supports VT-x and EPT technologies – Nehalem microarchitecture. $5^{th}$ generation CPU supports VT-x, EPT, and PT technologies – Broadwell microarchitecture.

In the next section we will present MemoryMonRWX, which is said to be free from all above mentioned drawbacks.

Table 2 Summary table of memory monitoring projects

| Title, year | Controlling the type of access | | | Supported OS | Required CPU Generation |
|---|---|---|---|---|---|
| | Read | Write | Execute | | |
| HIMA, 2009 | – | – | + | only Linux | $2^{nd}$ |
| HyperSleuth, 2010 | + | – | – | Windows 7, 8, x64 | $2^{nd}$ |
| CXPInspector, 2013 | – | – | + | Windows 7 x64 | $5^{th}$ |
| SPIDER, 2013 | + | + | – | only Linux | $2^{nd}$ |
| DRAKVUF, 2014 | – | – | + | Windows 7 x64 | $2^{nd}$ |
| HyperTap, 2014 | – | + | + | Windows XP, Vista, 7 | $2^{nd}$ |
| *MemoryMonRWX, 2017* | **+** | **+** | **+** | *Windows 7-10 x64* | *$2^{nd}$* |

## 3. DESIGN OF MEMORYMONRWX – THE MEMORY MONITOR HYPERVISOR

In this section we present the main new contributions of this paper. This part covers the proposed ideas of how to apply EPT technology to trap and manage memory access, and also the design of a proposed hypervisor-based memory monitoring system, which is able to simultaneously track all types of memory access: read, write, and execute. We have named it the MemoryMonRWX system – which stands for Memory Monitor of Read, Write, and eXecute access.This system registers read and write access to the specific range of virtual memory addresses as well as revealing executable code on these memory pages. This system supports multi-core processors and consequently runs itself on each core.

This is the first memory monitoring system which can trap even one-byte modification of guest OS, while all other solutions work on page granularity level. MemoryMonRWX did well when tested on the latest Windows 10 14393 x64 system.

### 3.1. How to Apply EPT-Technology to Trap and Control Memory Access

Modern malware and spyware rootkits apply the following typical techniques to protect themselves from being detected: malware can read and modify the content of data and code in the memory.

There are two main scenarios. First we can monitor memory access from a suspicious driver, which is loaded from known addresses range. In the second scenario access to the suspicious memory addresses is controlled. To cover these two cases MemoryMonRWX needs to monitor and control access by using source address range (SRC range) and destination address range (DST range). MemoryMonRWX skips accesses from all other ranges (OTH range). In this paper SRC range, DST range, and OTH range do not intersect and this sum (SRC range + DST range + OTH range) is in the entire virtual context. SRC range includes virtual addresses, access from which will be trapped and DST range includes virtual address, access to which will be trapped. MemoryMonRWX controls memory access only from the SRC range to DST range and skips all other accesses.

To provide such surveillance we apply a new Intel VT-x with Extended Page Table (EPT) technology, which significantly expands the existing bare-metal hypervisor facilities. Some details of EPT technology were given above in section 2.2.2. In a nutshell, by applying EPT paging structures, EPT technology provides a mechanism which can intercept and control access to the memory pages.

Using EPT structures we can implement the control of memory access from the SRC range to DST range and skip all other accesses, see Figure 6 *a*). Once guest memory is accessed, the translation between guest virtual address (GVA) to guest physical address (GPA) occurs, as shown in arrow (1) in Figure 6 below. After page walk is completed, the translation between GPA to host physical address (HPA) occurs, as shown in arrow (2). We receive HPA during page walk via EPT structures, which are used as an intermediary. If this memory access is allowed by the EPT, PT entry access bits, hypervisor does not take control. If this memory access is disallowed according to the EPT page table access bits, it involves EPT violation and causes VM exit, as shown by arrow (3). Now hypervisor is able to log and control access to this memory page via modification of EPT paging structures, see arrow (4). After that control goes to physical memory (5) and comes back to the guest, see arrows (6), (7), (8).

Let us consider two main scenarios of using EPT structures. Firstly, we set allowing attributes in the EPT structures and hypervisor does not trap anything, see Figure 6 *b*). Secondly, to intercept each read-access to the guest page we change the attributes on the corresponding EPT PT entry, see Figure 6 *c*).

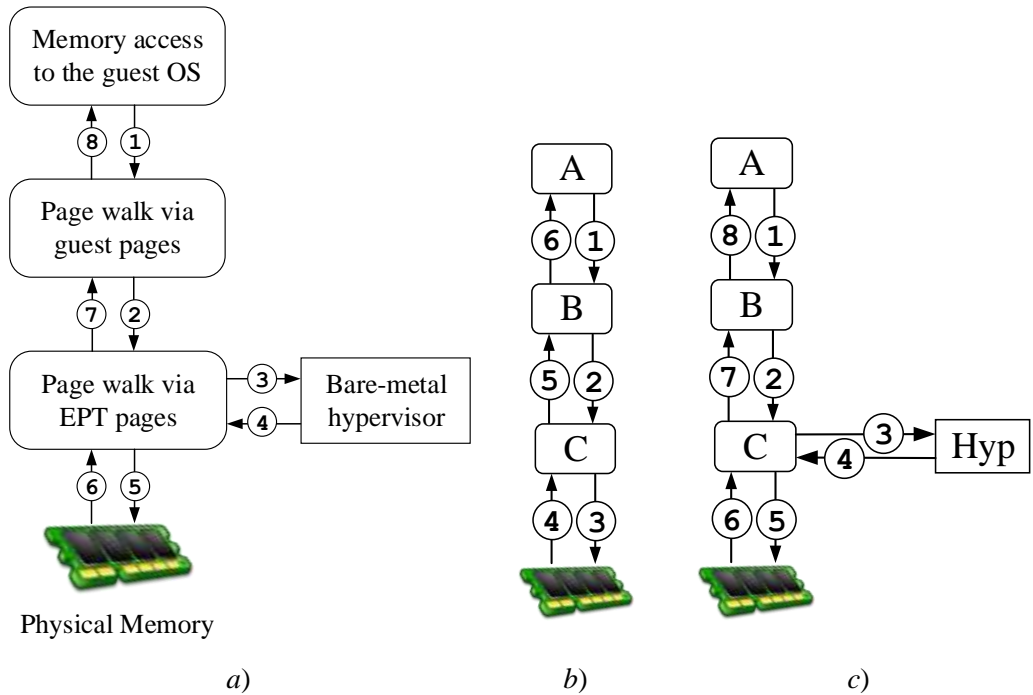

*a)*        *b)*        *c)*

Figure 6 Algorithm of intercepting memory access using EPT: a) General view;
b) EPT structures have allowed attributes; c) EPT structures have disallowed attributes

These two scenarios can be combined in the following way, which includes 5 steps. We control memory access from SRC range by resetting execute attributes in the corresponding EPT structures. Next we filter all these accesses to the DST range by resetting read- write- execute attributes in the EPT structures, which correspond to DST range.

**Step 1 (Trapping SRC range execution).** To separate only desired access from all other ones we use the following EPT structure as a trap, see Figure 7 *a)*.

This structure helps to intercept only execution access from SRC range, while all other accesses are skipped. This is an EPT normal view structure. In this structure DST range and OTH range will not be relevant. As a result, any code execution on SRC range involves EPT violation and causes VM Exit, and so we move on to Step 2.

**Step 2 (VM-Exit, because of execution on SRC).** To understand what this code is trying to achieve, we use the following EPT structure as a trap, see Figure 7 *b)*. Actually we can use only one EPT structure, but it requires updating each time during the change of the EPT pointer.

With this EPT structure the hypervisor will receive a VM Exit only if the code, which has been trapped on Step 1, is trying to access the DST range. This is in the EPT monitor view structure. Now any access to the DST range generates a VM Exit again and we move on to Step 3.

**Step 3 (VM-Exit, because of access to DST).** At this point, control goes to the hypervisor again. If we need to trap and monitor memory access, we log all related information: SCR address, DST address, type of access (read/write/execute), byte values some of which may have been read or overwritten. But if we need to protect sensitive data (or code) from being read or prevent important data (or code) from being overwritten we apply the following 3 procedures:

1. Change EPT.PFN value of the secret page to another one, for example, to the null page.

2. Allow access to this page by setting 'true' to EPT.DST.read and EPT.DST.write.
3. Set Monitor Trap Flag (MTF).

Setting this flag enables the system to generate VM Exit system after executing each instruction (Zhu, 2014). After guest OS reads the replaced page and executes just one instruction, the control goes to the hypervisor, because of VM Exit, and we move on to Step 4.

**Step 4 (VM-Exit, because of MTF).** By now we will have protected the secret data (or code) from being read and tampered with. To get ready to intercept a new memory access, we restore the configuration by applying the following 3 procedures:

1. Restore EPT.PFN value to the original one.
2. Restore permission of EPT.DST.read and EPT.DST.write by setting 'false' value.
3. Clear the MTF.

After that any access to the DST range will generate VM-Exit, and we move on to Step 3. Any execute access on OTH range will also generate VM-Exit, and we move on to Step 5.

**Step 5 (VM-Exit, because of execution on OTH).** Now we check if this VM Exit address belongs to the SRC range. If it does not, it means that this code is out of our control and so we do not have to control it. So we change EPT back from monitor view to normal view in order to be ready to trap a new code execution on SRC range, we move on to Step 1.

The interaction between these five steps is presented in Table 3 and Figure 8.

We checked the proposed idea of using EPT to control memory access by developing a MemoryMonRWX hypervisor, which is presented in the next section.

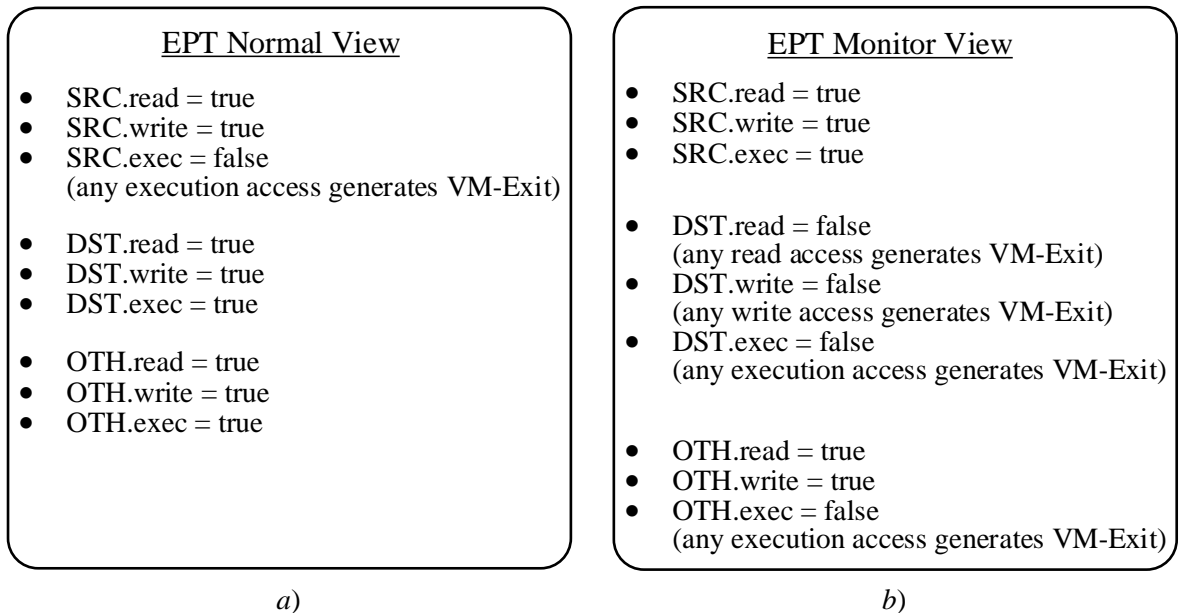

*a)*          *b)*

Figure 7 The content of EPT structures: a) EPT normal view, b) EPT monitor view

Table 3 Summary table of VM-Exit manipulations if access address belongs to SRC range

| Type of Access | Current Address | |
|---|---|---|
| | **Inside DST Range** | **Outside DST Range** |
| **Read / Write** | VM-exit & Recorded | Nothing |
| **Execution** | VM-exit & Recorded & Switch to Normal View | VM-exit & Switch to Normal View |

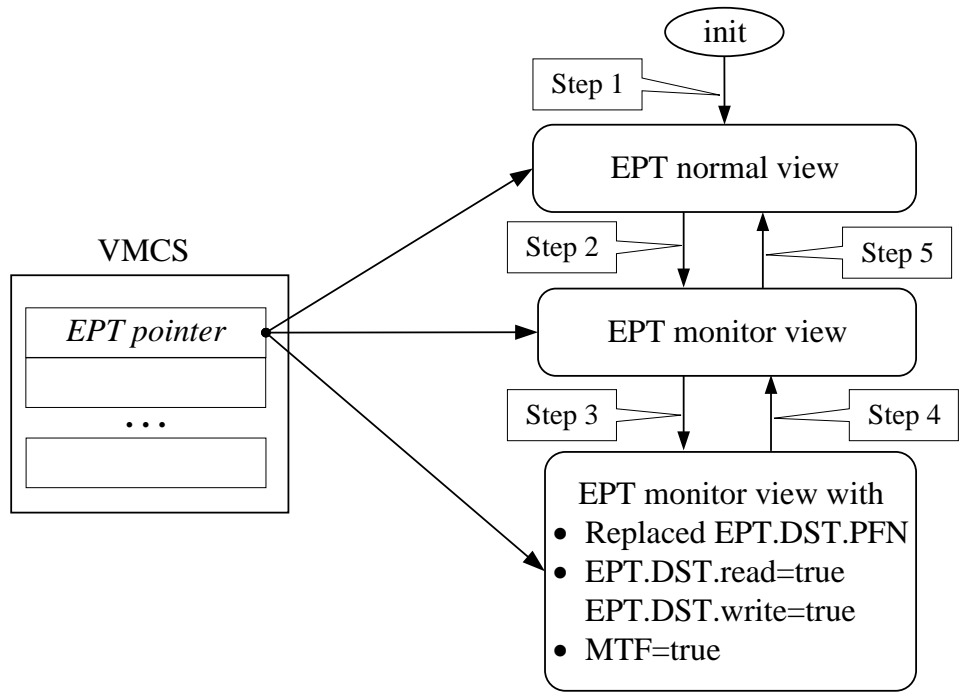

Figure 8 The proposed interaction between EPT views to log and control memory access

### 3.2. Architecture and Major Components

We have developed a hardware based hypervisor – MemoryMonRWX (Tanda, 2016-a), which leverages two Intel technologies: VT-x and Intel VT-x with EPT. MemoryMonRWX includes the following components: HyperPlatform, Image Load Detector, Source/Destination Range Manager (Src/Dst Range Manager), Virtual-to-Physical Map Manager (V2P Map Manager), and EPT controller.

A summary of the way this system works is shown in Figure 9. HyperPlatform is the main component of this system, which is a bare-metal hypervisor or virtual machine monitor (VMM). HyperPlatform is a minimal hypervisor, which is specifically designed for intercepting a variety of events in the guest OS and was firstly presented in REcon conference in 2016 (Tanda & Korkin, 2016).

After the MemoryMonRWX has been loaded, Image Load Detector forms a SRC/DST memory range of guest virtual memory addresses. Image Load Detector includes both type of ranges: pre-configured ranges, which include, for example, the addresses of critical memory areas and the addresses of recently loaded drivers, which are added automatically. In this situation, the addresses of recently loaded drivers are SRC addresses and DST addresses and these are critical memory areas. It is possible to specify your own set of SRT and DST ranges by modifying the code of MemoryMonRWX.

Src/Dst Range Manager takes requests from the Image Load Detector with SRC/DST virtual addresses ranges. This manager asks the EPT controller to update EPT settings for the stored ranges so that VM-Exit occurs when guest OS drivers from SRC ranges attempt to access any of the DST ranges.

V2P Map Manager maintains the mapping of virtual (VA) to physical addresses (PA). This manager takes requests for addition VAes from both SRC and DST ranges and stores them along with their corresponding PAes. Once any of following events occurs, HyperPlatform, requests V2P Map Manager to check whether any pair of VA:PA needs to be refreshed: translation Lookaside Buffer (TLB) flush; completion of #PF occurs due to access to the non-present page.

TLB flush indicates that any of previously valid VA:PA mapping via the page table entry has been changed, as for example, when the VA page is paged-out. The latter indicates that a new VA:PA mapping has just been established, for example, in case of paged-in page. V2P Map Manager will update the pair of VA:PA mapping in both cases.

EPT controller manipulates the guest OS behavior during the access to/from the configured memory regions. EPT controller is responsible for initializing and updating the EPT Paging Data Structures, handling EPT violation, and recording memory access. First, EPT controller accepts requests for updating the EPT setting from Src/Dst Range Manager for SRC and DST ranges. Second, EPT controller updates EPT Paging Data. Structures of a given range to trigger VM-Exit when this range is accessed. Third, EPT controller is notified by HyperPlatform, when VM-Exit has occurred via the mechanism of EPT violation. EPT controller checks whether the access should be logged by asking if the accessed VA is inside the DST ranges and the current code counter, for example, if the value of RIP register, is within the SRC range.

MemoryMonRWX provides fine-grained analysis by intercepting an access to the memory page. The logging process is only done when an EPT violation has occurred only on the configured address range. We do not log the accesses attempts to the EPT controlled pages, which do not belong to the configured ranges of memory addresses.

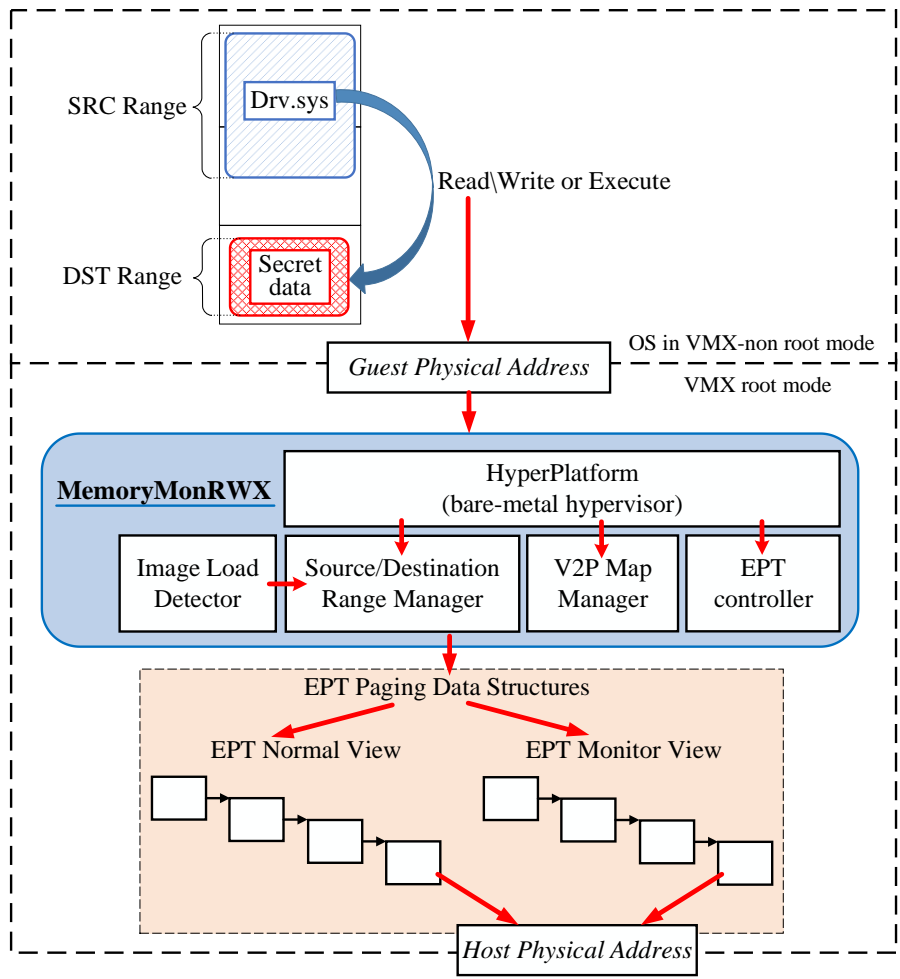

Figure 9 Architecture and Major Components of MemoryMonRWX

MemoryMonRWX traps access to the configured SRC and DST address ranges using two EPT paging structures: EPT normal view and EPT monitor view, see Figure 9. Note that each processor has those two structures so that multi-core systems can be supported. The normal view is used for the default state and the monitoring view is used when a guest is inside an SRC range. Details about the configuration and interaction between EPT normal view and EPT monitor view are given above in Section 3.1.

During processing MemoryMonRWX saves its log into the file C:\Windows\MemoryMon.log.

An example of this log is present in Figure 10 *a*). The first line indicates that a virtual address FFFFFA800194A468 is executed and its potential return address is FFFFF80002AD8C1C. Since execution of a non-image region is not always triggered by the CALL instruction, a reported return address can be wrong. For instance, the last line reports return address 0000000000000004. The return address is calculated in the following way. This address is the content of a memory address, specified by RSP at the point of EPT violation, ReturnAddr=*RSP. Actually, we do not know, execution on which this particular instruction has been trapped. To reveal the precise call stack we are planning to leverage the Intel Processor Trace (PT) mechanism.

To resolve symbol names in this log, a user-mode parser has been developed (Tanda, 2016-b). An example of a result log is presented in Figure 10 *b*).

MemoryMonRWX offers good compatibility with the all major Windows platforms. For instance, MemoryMonRWX supports and can monitor Windows 7, 8.1 and 10 on both x86 and x64 architectures with more than one core.

To ensure simplicity of its extension by researchers, MemoryMonRWX is designed to be small. As shown in Figure 11, it is made up of less than 12,000 lines of code, which is less than 3% of Xen, for example. Also, it can be compiled on Visual Studio without requiring any assistance from 3rd party libraries. MemoryMonRWX can be debugged with WinDbg just like a common Windows driver. Moreover, for rapid development, C++ and STL can be used if preferred.

We can conclude that the proposed MemoryMonRWX system has the following competitive advantages. First, it traps any accesses – read, write, and execute even to as little as one byte in the memory. It occurs due to leveraging EPT technology, which provides only page granularity level, and further processing, which reveals access even to one byte. Second, it supports multi-core processors via activating the VMX mode on each core. Finally, this system supports the newest Windows 10 14393 x64. Also, MemoryMonRWX can function as the basis for other cybersecurity solutions, for example, to monitor the activities of Device Driver Interfaces – DDIMon (Tanda, 2016-c), to detect unauthorized elevation of privilege – EoPMon (Tanda, 2016-d) as well as providing a mechanism to research and deactivate the PatchGuard – GuardMon (Tanda, 2016-e; Tanda, 2016-f, and Tanda, 2016-g).

```
[EXEC] *** VA = FFFFFA800194A468, PA = 000000007fe89468, Return = FFFFF80002AD8C1C, ReturnBase = FFFFF80002A5A000
[EXEC] *** VA = FFFFFA8003D46007, PA = 000000007db46007, Return = FFFFFA800194A4AD, ReturnBase = 0000000000000000
[EXEC] *** VA = FFFFFA8003D47580, PA = 000000007db47580, Return = FFFFFA8003D460B0, ReturnBase = 0000000000000000
[EXEC] *** VA = FFFFFA8003D4AE1C, PA = 000000007db4ae1c, Return = FFFFF80002AD7B69, ReturnBase = FFFFF80002A5A000
[EXEC] *** VA = FFFFFA8003D4856B, PA = 000000007db4856b, Return = 0000000000000004, ReturnBase = 0000000000000000
```
*a)*

```
executed fffffa800194a468, will return to fffff80002ad8c1c nt!KiRetireDpcList+0x1bc
executed fffffa8003d46007, will return to fffffa800194a4ad
executed fffffa8003d47580, will return to fffffa8003d460b0
executed fffffa8003d4ae1c, will return to fffff80002ad7b69 nt!ExpWorkerThread+0x111
executed fffffa8002626629, will return to                4
```
*b)*

Figure 10 Fragments of MemoryMonRWX log: a) raw data b) parsed data with resolved symbols names

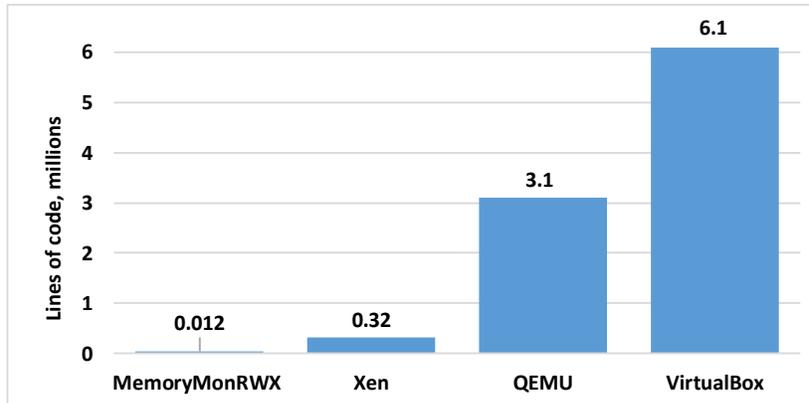

Figure 11 Comparison in lines of code of hypervisor projects,
MemoryMonRWX is made up of less than 12,000 lines of code, which is less than 3% of Xen

### 3.3. Three Demos of MemoryMonRWX

This sections covers three demonstrations of applying the MemoryMonRWX system. In the first example MemoryMonRWX stops the activity of a privilege escalation kernel mode exploit by detecting writing and causing BSOD. The second case demonstrates the read protection ability of MemoryMonRWX to prevent PatchGuard from being disabled. The final case deals with applying MemoryMonRWX to detect a suspicious code execution using Turla rootkit as an example.

### 3.3.1. Integrity Case – MemoryMonRWX Prevents Modifications of Code & Data

Typical kernel-mode rootkits hook functions through rewriting a code and unlink their structures by DKOM. The main scheme of these attacks is shown in Figure 12. These attacks are also known as Semantic Value Manipulation (SVM) attacks. These can mislead security tools by manipulating data values directly in the kernel data structures. Similar attacks are proposed by Prakash, et al., (2015).

The proposed MemoryMonRWE is able to detect and prevent such attacks. As an example we consider the CVE-2014-0816 kernel mode exploit (Tanda, 2016-h), which modifies its value in the HalDispatchTable[1]. To do it we predefined ranges in MemoryMonRWE; set address of HalDispatchTable[1] as a destination address. As a result, after loading the exploit MemoryMonRWE traps this modification and this is then able to stop the guest OS which prevents further exploitation. The video demonstration of this case is shown in Tanda (2016-i).

MemoryMonRWX can also be used to guarantee the integrity of critical kernel-mode sections, system lists, as well as the integrity of configured ranges to protect proprietary programs and their data. In this case MemoryMonRWX plays the role of future HyperGuard (Hyper Guard), which will probably replace the existing and vulnerable PatchGuard system (Ionescu, 2015; Chauhan, 2016).

### 3.3.2. Confidentiality Case – MemoryMonRWX Prevents Reading Data

Memory content includes much sensitive information: keystrokes, passwords, their hashes, private cryptographic keys, and even the fragments of decrypted data. Various rootkits attacks focus on kernel-level memory disclosure. The scheme of these attacks is given in Figure 13.

A description of the attacks of crypto key disclosure in the OpenSSH, Nginx server, and CryptoLoop is considered in Liu et al., (2015). A memory-based keylogger, which intercepts keystrokes by reading the content of DEVICE_EXTENSION of the kbdhid.sys driver, has been proposed by Ladakis et al., (2013).

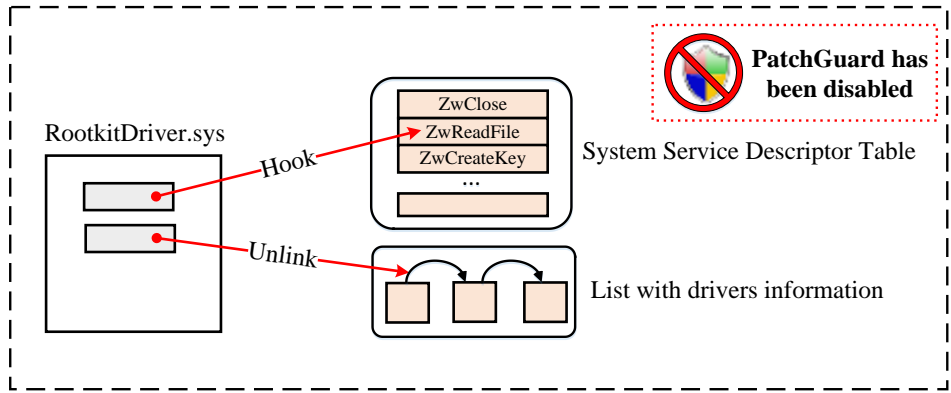

Figure 12 Code and data modifications attacks in the kernel-mode memory

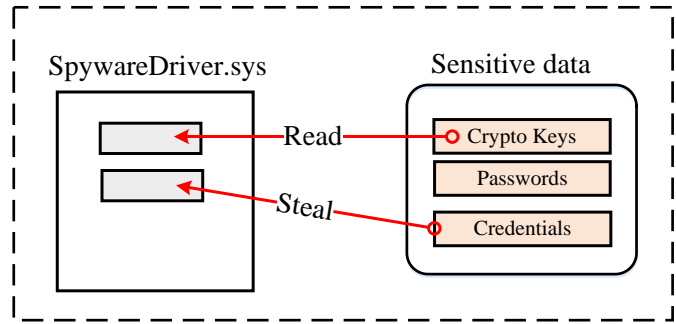

Figure 13 Spyware driver reads and steals sensitive data

To demonstrate that MemoryMonRWX has the ability to prevent read-access to sensitive data in the memory we use another kernel-mode exploit "Disarms PatchGuard (DisPG)", as has been proposed by Tanda (2016-j). One of the components of DisPG reads the value from nt!PoolBigPageTableSize, which stores the address of the big page pool table Sylve, et al., (2016). To prevent such unauthorized reading we predefined the MemoryMonRWX destination range using the address of nt!PoolBigPageTableSize and also changed the logic of intercepting in the following way: Thus any unauthorized reading access attempts to this memory content will be redirected to a fake zero page. As a result, DisPG reads the replaced fake zero value and fails to disable PatchGuard. Video demonstration of this case is loaded in (Tanda, 2016-k).

Thus MemoryMonRWE prevents any unauthorized reading access of the sensitive data.

### 3.3.3. Real World Case – Applying MemoryMonRWX to Analyze Turla Rootkit

Another rootkit technique moves malware executable code outside the driver section. As a result, kernel memory includes unknown pages with an executable code.

MemoryMonRWX is able to reveal such executable code as well as providing facilities to analyze it with the help of a disassembler. To demonstrate these facilities, we use Turla rootkit (also known as Uroburos rootkit).

We tested MemoryMonRWX on the 64bit version of Windows 7 against the Turla rootkit and confirmed that MemoryMonRWX is able to detect execution of non-paged pool and that the executed region contained unpacked rootkit code (Tanda, 2016-l).

### 3.4. Benchmarks

Performance measurement was conducted on the 64bit version of Windows 10 running on a Macbook Air with Intel Core i7-4650U, 8GB RAM and SSD flash storage. In this experiment, we executed Novabench (Novabench, n.d.) and PCMark8 Home (PCMark8, n.d.) on the system with and without MemoryMonRWX. Compared overhead in ratio is shown in Figure 14. We can measure how much the system performance changed in comparison to 0%, which indicates the system operating without those hypervisor tools.

The results showed that performance degradation kept to less than 10% in all tests except the Novabench Graphics Tests. We surmise that the reason for the higher overhead on this test is caused by frequent TLB flush led by active memory access, yet this has not been investigated so far. Users should experience much less overheads during their routine work: opening and saving documents or surfing the Internet.

### 4. CONCLUSIONS & FUTURE WORK

In this paper we have achieved the following results:

1) We are able to reveal and prevent malicious activity by logging and controlling read-, write-, and execute-memory access in a real time mode.
2) We developed a MemoryMonRWX hypervisor to ensure the integrity and confidentiality of both code and data. This system helps to detect kernel-mode malware, even if this malware applies popular OS-based prevention techniques.
3) MemoryMonRWX can be used to monitor access to the memory for a variety of different purposes: driver tracking, reverse engineering, detection of unknown malware, verification and protection of proprietary software.
4) We demonstrate that MemoryMonRWX can be used in practice, the evaluation of its benchmarks shows that its degradation is quite small.
5) MemoryMonRWX is a tiny open-source project which can be easily used by students and post-graduate students during their research activity.

With regard to future work we would like to suggest the following ideas

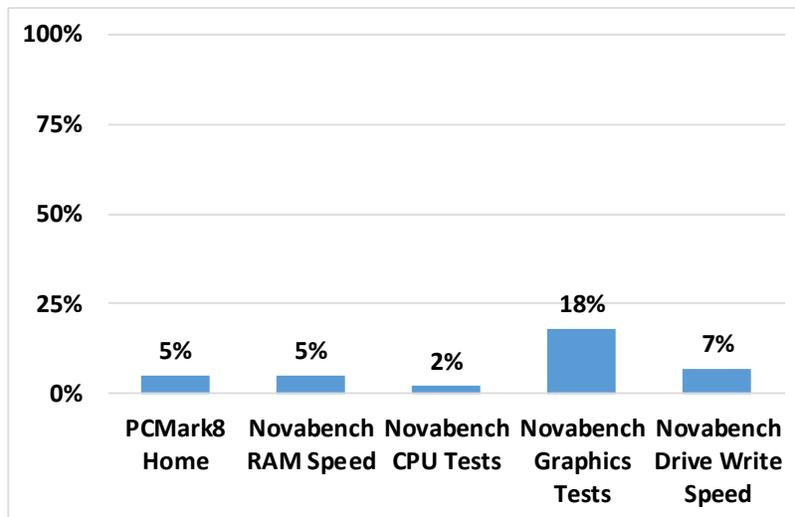

Figure 14 MemoryMonRWX overhead
0% indicates the system without hypervisor, 100% – full system overload

## 4.1. Hypervisor-based Warden Controls Access to The Memory

We propose an idea of how to improve PatchGuard facilities and make it more resilient. We propose an idea for hypervisor-based warden (HyperWarden) which is not vulnerable to all kernel-mode malware manipulations, because it runs in a more privileged mode. Existing PatchGuard provides integrity for Windows kernel code and detects unlinking attacks on the structures from process and drivers lists (Ionescu, 2015). It does not protect the integrity of the full content of these structures in the memory (Ch40zz, 2015) as well as not providing any mechanisms to protect memory of the 3$^{rd}$ party drivers from being tampered with.

HyperWarden will exclude all these drawbacks. It will provide flexible protection for all data in the memory using MemoryMonRWX as its basis. By dynamic configuration of SRC/DST ranges and allowed types of memory access we can guarantee the data and code security. HyperWarden will avoid any modification of critical Windows code and multiple structures in the memory. It will allow modification and read critical Windows data only through the Windows kernel code. To protect the integrity and confidentiality of the code and data of the third part drivers HyperWarden will provide API to configure regions of memory, which need be protected. HyperWarden will support functions to activate/deactivate memory protection as well as adding\deleting protected memory areas for each driver. As a result, HyperWarden helps to provide complex memory security: protect integrity for OS critical areas as well as integrity and confidentiality of users for the configured memory areas.

## 4.2. Protection of Cloud Computing Systems

One of the possible scenarios of large scale application of MemoryMonRWX is to protect Cloud Computing Systems from being tampered by exploits and malware (Murakami, 2014). Private Cloud Computing Systems such as Amazon, Google, and Microsoft provide their clients with common services, whose behaviors are little altered. For each specific Software as a Service (SaaS) we can generate various behavior signatures, which correspond to typical operations with memory, and in this way avoid the leakage of users' data.

## 4.3. Visualize Memory Access

Another suggestion is to visualize registered memory access using various techniques. The idea is to create a Dynamic Memory Map, demonstrating which driver or code has access to specific data in the memory. It may also monitor the frequency, amount of accessed data, and the content of memory. The first step is to draw a Static Memory Map with loaded drivers together with the allocated data using rectangles. It will look like a typical memory dump. The second step is to trap the access from each driver to the memory and then draw the corresponding arrow between the two blocks. The third step is to continue updating the picture and as a result this will show the Dynamic Memory Map. For first step we can use various data visualization techniques (ISOVIS, n.d.) and for the second step we can apply ideas from Rgat roject (Catlin, 2016).

## 4.4. Apply Raspberry Pi to Acquire Physical Memory Dump & Detect Hidden Software

We propose an idea of using modern IoT platforms such as a Raspberry Pi for the protection of computers and incident response. The idea is to expand the opportunities of CaptureGUARD by WindowsSCOPE (WindowsSCOPE, n.d.), which is only able to acquire the physical memory dump, and so significantly decrease the price of a new hardware platform.

First of all, we can use Raspberry Pi to acquire the dump of physical memory using the ExpressCard slot for PC and Thunderbolt interface for Mac. A tecnhique of dumping memory by FPGA on a PCMCIA card or ExpressCard slot was proposed by Aumaitre, and Devine, (2010). We are planning to apply an Inception software tool, which exploits PCI-based DMA. This tool can attack over FireWire, Thunderbolt, ExpressCard, PC Card and any other PCI/PCIe interfaces (Maartmann-Moe, n.d). After dumping we can use Raspberry Pi facilities to process this memory dump using Rekall Memory Forensic or Volatility Frameworks. We can also update detection software using a wireless connection, that is built-in to Raspberry Pi. This detection platform will also be resilient to malware attacks, because users do not work on it.

### 4.5. Implantable Medical Devices as a Target of Cyber Attacks

Another idea is to protect wireless Implantable Medical Devices (IMD) from being hijacked using remote control. The livelihood and welfare of patients ultimately depends on the precise work of these devices. Their work can be breached remotely by an intruder through using a wireless connection and thus can result in human losses. We propose the following action plan to protect IMD and make forensic investigation easier. We can maintain confidentiality, integrity, and authenticity of data by applying lightweight cryptography to a secure channel. We will suggest an intrusion detection system (IDS) to protect IMD from DoS attacks by disabling its input dispatcher temporarily (Darji & Trivedi 2013). IDS will protect battery IMD from being drained. We will describe the event logging system which is able to perform the forensic analysis in case of an incident occurring. Finally, we are planning to verify embedded software to reveal the vast majority of vulnerabilities.

As the first step, we will create an analogue of OneTouch Ping Glucose Management System & Insulin Pump by J&J, which was attacked recently (Finkle, 2016). We will apply Contiki OS for programming TI MSP430 microcontroller, which is used in these pumps. We demonstrate the vulnerability of this radio channel by unauthorized control and access to this pump. We will develop a complex cyber-security system, which will protect this IMD from being tampered with remotely, or from stealing data and draining the IMD battery by wireless DoS attacks.

### 5. ACKNOWLEDGEMENTS

We thank the anonymous reviewers for their constructive feedback to this work.

We wish to express our gratitude to Ashlyn King, an intern at Russian Flagship Center, University of Wisconsin–Madison, Madison, Wisconsin, US for her comments on the manuscript and equally helpful advice. Her voluntary contribution in reviewing this paper significantly improved its quality and timeliness.

We would like to thank Sarah Krueger, a teacher of English, Kenosha, Wisconsin, US for her time and effort in checking a preliminary version of this paper.

We would also like to thank Ben Stein, teacher of English, Kings Education, London, UK for his invaluable corrections of the paper.